\PassOptionsToPackage{desactivate}{linenoaa}
\newif\ifcolumnwiselinenumbers
\let\runningpagewiselinenumbers\relax

 \documentclass{aa}   %  2 column "printer" version

\usepackage{graphicx}
\usepackage{txfonts}
 \usepackage{natbib}  % OLI AKTIIVI V 2023 + 2024 PAPRUISSA
\usepackage{placeins}           % useful with \FloatBarrier, to keep 
\begin{document}

\title{Relation between sunspots and solar EUV irradiance changes during the Gleissberg cycle}

% \subtitle{Subtitle}

%%%%%%%%%%%%%%%%%%%%%%%%%%%%%%%%%%%%%%%%
% Please do not include ORCIDs next to author names.
% Only ORCIDs authenticated by individual authors in EDP Sciences editorial system will be taken into account.
% ORCIDs included here will be removed.
%%%%%%%%%%%%%%%%%%%%%%%%%%%%%%%%%%%%%%%%

%   \author{N. Puyaubreau\inst{1}
%        \and F. Déliat\inst{2}\fnmsep\thanks{Shows the usage of elements in the author field}
%        }

    \author{K. Mursula \inst{1}
      }

   \institute{Space Climate Group, Space Physics and Astronomy Res. Unit, University of Oulu,
             90014 Oulu, Finland\\
              \email{kalevi.mursula@oulu.fi}
 %             \email{\href{mailto:kalevi.mursula@oulu.fi}{kalevi.mursula@oulu.fi}}
    }

%   \institute{EDP Sciences, 91940, Les Ulis, France\\
%             \email{support@nestor-edp.org}
%             \thanks{Shows the usage of elements in the author field}
%            \and EDP Sciences, 91940, Les Ulis, France\\ }

%  \date{Received September 30, 20XX}

 \titlerunning{Sunspot - EUV long-term relation}
\authorrunning{Mursula}

%%%%%%%%%%%%%%%%%%%%%%%%%%%%%%%%%%%%%%%%%%%%%%%%%%%%%%%%%%%%%%
  \abstract
%%%%%%%%%%%%%%%%%%%%%%%%%%%%%%%%%%%%%%%%%%%%%%%%%%%%%%%%%%%%%%
% \abstract{}{}{}{}{}
% 5 {} token are mandatory
%   {Optional, leave empty if necessary.  The heading “Context” is used when needed to
%give background information on the research conducted in the paper}
%  % aims heading (mandatory)
%   {Mandatory. The objectives of the paper are defined here.} 
%  % methods heading (mandatory)
%   {Mandatory. The methods of the investigation are outlined here}
%  % results heading (mandatory)
%   {Mandatory. The results are summarized here.}
%  % conclusions heading (optional), leave it empty if necessary
%   {Optional, leave empty if necessary.  “Conclusions” can be used to
%explicit the general conclusions that can be drawn from the paper.}
{
Sunspots are the standard measure of solar magnetic activity, which are also used to estimate solar spectral irradiance over centennial time scales.
However, because of the lack of homogeneous, century-long spectral measurements, the long-term relation of sunspots and spectral irradiance has not been independently validated.
}
{
Here we aim to study the relation between sunspots and solar extreme ultra-violet (EUV) irradiance during the last 130 years, over the latest Gleissberg cycle, also called the Modern Maximum, when sunspot cycle heights varied by a factor of 2.5.
} 
{
We calculate the daily variation of the geomagnetic declination at six reliable, long-running stations, whose amplitude (or range) can be used as a centennial proxy of solar EUV irradiance.
We also compare this geomagnetic proxy to the solar MgII index of EUV irradiance over the 40-year interval of overlap.
}
{
We find that sunspot activity dominated over EUV irradiance when cycle heights increased in the early 20th century during the growth and maximum of the Modern Maximum, but EUV irradiance dominated over sunspots during the decay of the MM, when cycle heights decreased in the late 1900s. 
}
{
Our results suggest that the spot-facula ratio varies during Gleissberg cycle -type large oscillations of solar/stellar activity.
This modifies the estimated stellar evolution of the relation between brightness and chromospheric activity of the Sun and Sun-like stars.
}


   \maketitle

%%%%%%%%%%%%%%%%%%%%%%%%%%%%%%%%%%%%%%%%%%%%%%%%%%%%%%%%%%%%%%
\section{Introduction}
	\label{sec:Introduction}
%%%%%%%%%%%%%%%%%%%%%%%%%%%%%%%%%%%%%%%%%%%%%%%%%%%%%%%%%%%%%%

The height of sunspot cycles has varied dramatically during the roughly 400 years of telescopic monitoring.
Soon after the start of observations, sunspots almost disappeared in the second half of the 17th century in a period now called the Maunder minimum \citep{Spoerer_1887, Maunder_Spoerer_MNRAS_1890, Eddy_1976, Usoskin_2015}.
%, actually found by G. Spörer.
Since the restart of cyclic activity at the beginning of the 18th century, cycle heights have varied at a near-centennial Gleissberg cyclicity \citep{Gleissberg_1939, Ogurtsov_2002, Feynman_2014, Hathaway_LRSP_2015}.

The latest Gleissberg cycle, also called the Modern Maximum, started with low cycles at the turn of the 19th and 20th centuries, peaked at an all-time maximum during cycle 19 in 1950s \citep{Solanki_2004}, and returned back to a low level during cycle 24 in 2010s. 
Sunspot cycle heights experienced a variation by a factor of about 2.5 during this centennial variation of solar activity.
The ongoing cycle 25 is already more active than cycle 24 (for predictions, see, e.g., \cite{Pesnell_2020, Bhowmik_2023, Upton_2023}; for the current status of cycle 25, see https://www.sidc.be/SILSO/), which may indicate that the next Gleissberg cycle has already started.

Such a large long-term variation in solar activity has dramatically modified space weather conditions on the Earth \citep{Lockwood_SWSC_2017, Chapman_2020, Mursula_JGR_2022} and in the whole heliosphere during the last 100 years.
Although the brevity of space age limits direct solar spectral observations to the last 50 years, 
% the period when cycle heights were decreasing, 
indirect estimates based on various proxies 
% like the open solar flux 
suggest a significant centennial change even in solar spectral irradiance \citep{Ermolli_2013, Penza_2024}.
However, a detailed study of how closely the sunspots and spectral irradiance parameters follow each other
% more generally, how synchronously the different solar parameters vary
over the Modern Maximum, is still missing.

Several studies have found that the mutual relation between sunspots and solar 10.7\,cm radio flux changed from 1970s until 2010s so that the Sun emitted more of radio flux relative to sunspots in the last two decades of this period than in the earlier decades \citep{Tapping_2011, Tapping_2017, Bruevich_2019, Lastovicka_SpW_2023, Mursula_AA_2024}.
Since the solar 10.7\,cm radio flux is a close proxy of solar EUV irradiance \citep{Tapping_2013, Schonfeld_2019},
%, especially for monthly and longer time scales, 
this implies that sunspot activity decreased relatively faster than EUV irradiance when solar activity was weakening during the decay of the Modern Maximum.
This is also verified by direct measurements of the solar MgII core-to-wing ratio index \citep{Mursula_AA_2024}, which is a standard measure of solar EUV irradiance.

Already 300 years ago G. Graham found that the declination of the Earth's magnetic field
% of the geomagnetic field 
experiences a systematic daily variation with a maximum in the morning and minimum in the afternoon \citep{Graham_1724}.
In the early 1850s J. von Lamont \citep{Lamont_1851}, 
%E. Sabine, 
J.-A. Gautier \citep{Gautier_1852} and R. Wolf \citep{Wolf_1852} noted that the amplitude of the daily variation of declination follows the sunspot cycle, which was found somewhat earlier by S.H. Schwabe \citep{Schwabe_1844}.
Soon thereafter, when constructing his renowned sunspot numbers, Wolf noted that there is a highly significant linear relation between the yearly sunspot number and the yearly averaged declination amplitude \citep{Wolf_1859}.
He also used this relation to predict the declination amplitude from sunspots and, vice versa, to fill in data gaps in his sunspot series using the declination amplitude.

The basic ingredients of solar influence on the Earth's upper atmosphere were theorised early, but their observational verification took a longer time. 
\cite{Stewart_1882} suggested that the upper atmosphere is electrically conducting and a site of large, persistent electric currents, %which can cause magnetic effects on the ground and produce a diurnally varying declination.
and \cite{Schuster_1908} proposed that these currents are produced by solar extreme ultraviolet (EUV) radiation which ionises the dayside upper atmosphere.
However, satellite observations were needed to prove that solar EUV radiation indeed varies in phase with sunspot cycle \citep{Hinteregger_1979, Donnelly_1986}.
This gave the motivation for Wolf's linear correlation between sunspots and declination amplitude, more than a century after the initial finding.

Here we use the daily range of the East (Y)-component of the geomagnetic field at six long-running mid-latitude stations to develop a centennial proxy of solar EUV irradiance and to study its relation with sunspots over the whole Modern Maximum.
The paper is organised as follows. 
We introduce the data in Section \ref{sec:Data}, and derive the daily range (rY index) of the Y-component in Section \ref{sec:rY}.
Section \ref{sec:6st_mean_rY_SSN} studies the correlation between the 6-station mean rY index with sunspots, and Section \ref{sec:rY_earlier} extends this study to earlier decades using the mean rY of the four longest-running stations and the rY index of the longest Niemegk station.
In Section \ref{sec:Discussion} we discuss and interpret the obtained results in view of solar and stellar evolution, and give our conclusions.

%%%%%%%%%%%%%%%%%%%%%%%%%%%%%%%%%%%%%%%%%%%%%%%%%%%%%%%
% DATA AND METHODS
\section{Data and Methods}
\label{sec:Data}
%%%%%%%%%%%%%%%%%%%%%%%%%%%%%%%%%%%%%%%%%%%%%%%%%%%%%%%

We use magnetic observations from six long-operating standard stations: Hermanus (HER), Honolulu (HON), Kakioka (KAK), Niemegk (NGK), San Juan (SJG) and Tucson (TUC) (for coordinates and start years, see Table \ref{table:Stations}).
Data of four stations (HER, HON, KAK, SJG) are also used to calculate the storm-time Dst index \citep{Sugiura_1964} %Sugiura_1991
and its extension, the Dxt index \citep{Karinen_Mursula_2005}.
NGK is the base station for the Kp/Ap geomagnetic activity indices \citep{Matzka_2021}.
The common time interval for all the six stations starts in 1932, but observations in NGK started in 1890, i.e., 40 years earlier. 

We use three long-term measures of solar activity, the Wolf/International sunspot number (SSN), the group sunspot number (GSN) and 
%sunspot areas and two solar radio fluxes at slightly different wavelengths, the 10.7\,cm flux (F10.7) and the 30\,cm flux (F30).
the solar MgII index. 
Sunspots give a view of the evolution of photospheric magnetic activity, while the MgII index is the standard measure of solar EUV irradiance and a measure of activity 
%at two different altitudes of the chromosphere and lower corona. 
in the chromosphere and lower corona. 
For SSN we use the yearly sunspot number of version 2 \citep{Clette_2015, Clette_Preface_2016} and for GSN the collection of sunspot groups from 1610 to 2010 by \cite{Vaquero_2016}, a revision of the original collection by \cite{Hoyt_SP2_1998}.
However, we note that our results remain the same for SSN version 1, as well as other GSN collections.

The core-to-wing ratio of the Magnesium-II doublet at 280\,nm (MgII index) is a standard measure for solar UV-EUV irradiance \citep[see, e.g.,][]{DeLand_1993, Viereck_1999, Viereck_2001, Viereck_2004, Snow_2014}.
%(for a recent review see, e.g., Clette et al. \citeyear{Clette_2014}.
Magnesium-II measurements from several satellites were constructed to a long-term Mg II index, the so-called Bremen MgII composite index
%\footnote{$\text{https://www.iup.uni-bremen.de/gome/gomemgii.html}$} 
updated by M. Weber (https://www.iup. uni-bremen.de/gome/gomemgii.html). 
%($https://www.iup.uni-bremen.de/gome/gomemgii.html$). 
The MgII index is an index of the overall chromospheric activity and, as will be discussed later in more detail, has a close connection to solar plages.

%%%%%%%%%%%%%%%%%%%%%%%%%%%%%%%%%%%%%%%%%%%%%%%%%%%%%%%
% rY
\section{Daily range of the Y-component}
\label{sec:rY}
%%%%%%%%%%%%%%%%%%%%%%%%%%%%%%%%%%%%%%%%%%%%%%%%%%%%%%%

The ionisation of the Earth's dayside upper atmosphere by solar EUV irradiance, and the Earth's rotation form the $S_q$ current system \citep[for a review, see][]{Yamazaki_2017}.
The $S_q$ current system consists of two current vortices (see the left panel of Fig. \ref{fig:S1}), one in each hemisphere, whose turning points in the morning and afternoon deflect the East (Y)-component to the east and west, respectively, leading to a regular daily variation of declination.
Depicting the Y-component in local time yields a closely similar variation for all the five northern hemisphere stations with a morning maximum and afternoon minimum, while the southern station (HER) shows an opposite variation (see Fig. \ref{fig:S1}).

\begin{figure*}[h]  %% 
   \centering
    \includegraphics[width=0.95\linewidth]{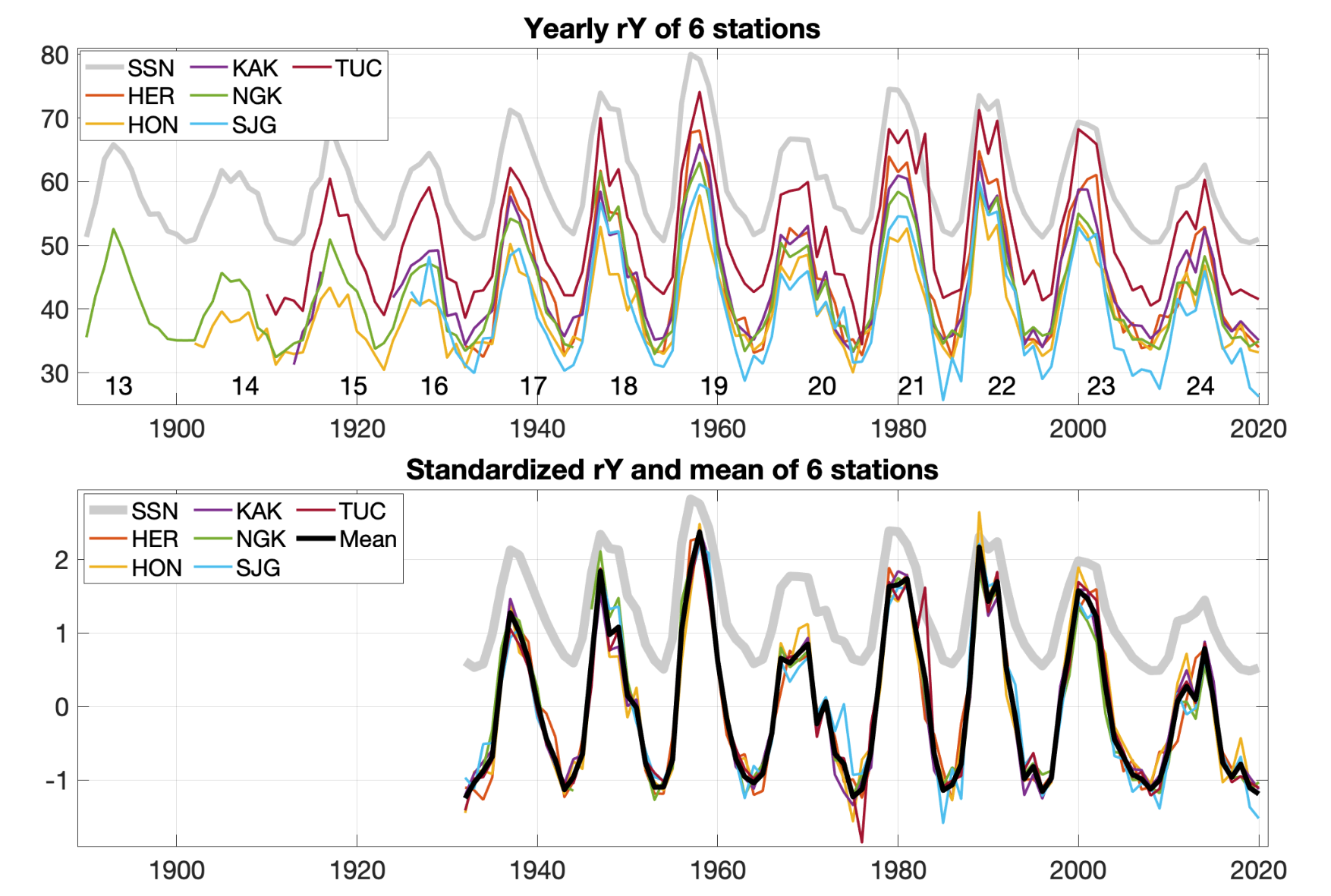}
    \caption{Top: Yearly rY indices of the six stations over their respective time intervals. 
    Cycle numbers are indicated at the bottom of the panel.
    Bottom: standardized yearly rY indices of the six stations and their mean for the common time 1932-2020. 
 Sunspots (out of scale) are depicted as grey thick line in both panels for comparison.} 
   \label{fig:figure_rY_6st_abs_st_mean}
\end{figure*}

The difference between the daily maximum and minimum of the Y-component, i.e., the range of daily variation is called the rY index \citep{Svalgaard_EUV_2016}. 
% (Wolf?, Svalgaard, Mursula).
We have calculated the yearly averaged rY indices for the six stations and depicted them in the top panel of Fig. \ref{fig:figure_rY_6st_abs_st_mean}.
(We use all days since the daily variation is practically the same for quiet and all days, as first noted by \cite{Chree_1913}).
% \cite{Chree_1913} first observed that the relation between yearly declination amplitude and sunspots was the same for all days and quiet days.
The rY indices of all stations depict a cyclic evolution, following the sunspot cycles and even reproducing the centennial variation of cycle heights over the Modern Maximum, with the highest rY cycle found during the highest sunspot cycle 19.

%%%%%%%%%% FIGURE 1 %%%%%%%%%%%%

Figure \ref{fig:figure_rY_6st_abs_st_mean} shows that the rY cycle amplitudes and even the absolute scales of the six rY indices are quite similar although the absolute level and the secular evolution of the Y-component is very different at the six stations. 
(See Fig. \ref{fig:S2} for the Y-component and the daily maxima and minima at the six stations). 
% Only TUC remains somewhat higher and SJG somewhat lower than the other stations.
In order to set all the stations to the same level we have calculated the standardized yearly rY indices for all stations in 1932-2020, the common time interval.
(standardization removes the mean of rY series and divides by its standard deviation.)
The six standardized yearly rY indices are shown in the bottom panel of Fig. \ref{fig:figure_rY_6st_abs_st_mean} together with their 6-station mean.
Although some stations, in particular HON, SJG and TUC, occasionally deviate from the mean curve, none of the six stations deviate systematically from the mean rY curve.
Since standardization does not affect the trend of the index, this means that the rY indices of all the six stations, as well as their mean, evolve closely similarly during the time interval after 1932.

%%%%%%%%%%%%%%%%%%%%%%%%%%%%%%%%%%%%%%%%%%%%%%%%%%%%%%%
%  6-station mean rY
\section{Correlating 6-station mean rY and sunspot numbers in 1932-2020}
%\section{Correlating 6-station mean rY and sunspot numbers}
\label{sec:6st_mean_rY_SSN}
%%%%%%%%%%%%%%%%%%%%%%%%%%%%%%%%%%%%%%%%%%%%%%%%%%%%%%%

\begin{figure*}[h]  %% 
   \centering
  \includegraphics[width=0.98\linewidth]{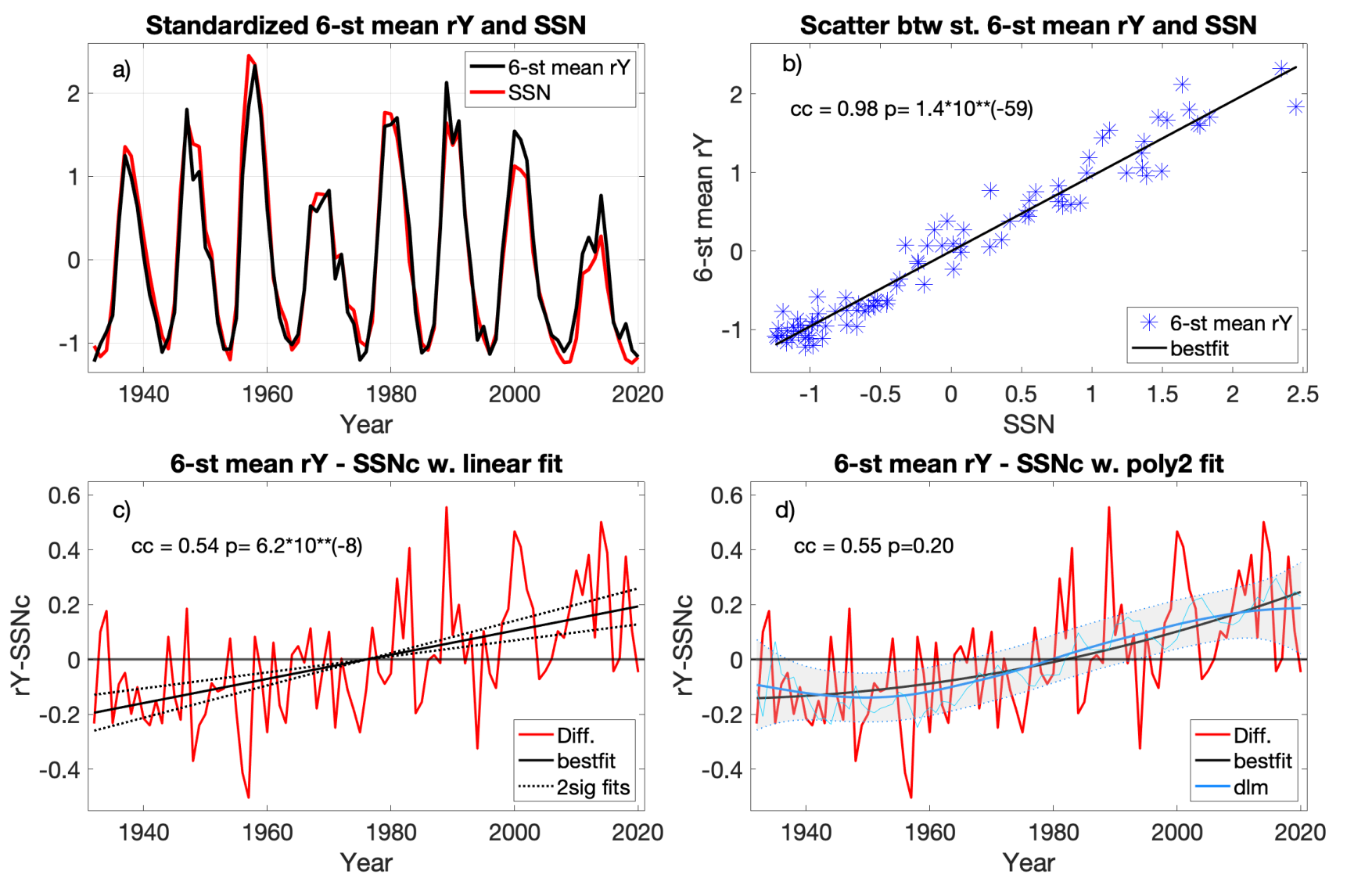}
    \caption{a) standardized yearly 6-station mean rY index (black) and sunspot number (red). 
    b) Scatterplot of standardized 6-st mean rY index and SSN (blue stars), the best-fit line (black) and its correlation coefficient and p-value. 
    c) Difference of standardized 6-st mean rY and correlated SSN.  
   Best-fit line (solid) with correlation coefficient and p-value, as well as lines with slopes that are two standard deviations above or below the best-fit line slope (dotted). 
   d) Same difference of standardized 6-st mean rY and correlated SSN as in panel c, with best-fit second-order polynomial fit (black) and the dynamic linear model curve (blue) with 95\% errors (grey area).  } 
   \label{fig:figure_rY_6st_mean_SSN_4panels}
\end{figure*}

We have standardized the sunspot number (SSN) over the same time interval and depicted it together with the 6-station (6-st) mean rY index in Fig. \ref{fig:figure_rY_6st_mean_SSN_4panels}a. 
One can see that the cycles of standardized rY and SSN have roughly equal amplitudes.
This means that a change of one unit of standard deviation in sunspot activity (about 72) produces a change of roughly one unit of standard deviation in the mean rY.

%%%%%%%%%% FIGURE 2 %%%%%%%%%%%%

The scatterplot of Fig. \ref{fig:figure_rY_6st_mean_SSN_4panels}b shows an excellent linear correlation between the mean rY index and sunspot number.
Correlation coefficient is 0.98, which implies that 96\% of the variation of yearly mean rY is explained by sunspot variability.
%, i.e., by solar activity.
The p-value is extremely small ($1.4\cdot 10^{-59}$).
Note also that the intercept of the best-fit line is practically zero (-0.0012) and the slope is 0.96, in agreement with the estimated explaining capability.
 
Despite this excellent correlation, a detailed analysis of Fig. \ref{fig:figure_rY_6st_mean_SSN_4panels}a reveals a small but systematic difference in the long-term trend between the mean rY and SSN.
Most maxima of the first sunspot cycles are slightly higher than the corresponding peaks of the mean rY index.
Even more clearly, the opposite is the case for the last three solar cycles, where rY cycle peaks are higher than sunspot peaks.
Moreover, the two minima around cycle 24 are lower in sunspots than in the rY index.

%The relation between the mean rY index with respect to the sunspot number is studied in detail in Figs. 
Figures \ref{fig:figure_rY_6st_mean_SSN_4panels}c and \ref{fig:figure_rY_6st_mean_SSN_4panels}d depict the difference (residual) between the standardized mean rY index and the correlated standardized sunspot number obtained from the correlation of Fig. \ref{fig:figure_rY_6st_mean_SSN_4panels}b.
This difference is, on an average, somewhat negative from 1930s until 1970s and clearly positive from 1980s onward.
Accordingly, the mean rY index has an increasing trend with respect to sunspot number during the depicted time interval, as already seen in panel a.
Figure \ref{fig:figure_rY_6st_mean_SSN_4panels}c includes a best-fit line to the rY-SSN difference. 
Despite the rather large scatter of individual yearly differences, the linear fit is highly significant (p=$6.2\cdot 10^{-8}$) and explains almost 30\% of the variation (cc = 0.54).
The slope of the best-fit line ($0.0044 \pm 0.0007$) differs from zero very significantly, which is also depicted by the two dotted lines in Fig. \ref{fig:figure_rY_6st_mean_SSN_4panels}c whose slopes deviate from the best-fit line slope by two standard deviations.

Despite the success of the linear fit, the temporal evolution of the rY-SSN difference may not be strictly linear in time over the whole time interval included.
%The linear fit of Fig. \ref{fig:figure_rY_6st_mean_SSN_4panels}c may not be the best-fitting low-order polynomial to the rY-SSN difference. 
The slightly nonlinear nature of this evolution is suggested, e.g., by the fact that the mean value of this difference (about -0.13) in the first three decades (1930s-1950s) remains fairly constant, as does the mean difference (about 0.19) of the last two decades.
In fact, most of the increase in this difference takes place between 1970s and 2000s.

Figure \ref{fig:figure_rY_6st_mean_SSN_4panels}d shows the same difference as panel c, with a fit to a second order (quadratic) polynomial, which gives almost the same correlation coefficient (cc=0.55) as the linear fit.
However, the second order term of this polynomial is not significantly different from zero (p = 0.20). 
We have also included in Fig. \ref{fig:figure_rY_6st_mean_SSN_4panels}d the local trend estimate (blue line) and its 95\% error (shaded area) using the dynamic linear model (dlm; \cite{Laine_2014}), which allows the regression coefficient to vary in time.
% and a smaller root-mean square error (RMSE=0.1763 vs 0.1781 for linear fit) 
The dlm trend flattens at either end, but the time interval is too short to give significant evidence for a possibly nonlinear evolution.

%%%%%%%%%%%%%%%%%%%%%%%%%%%%%%%%%%%%%%%%%%%%%%%%%%%%%%%
\section{Extending rY-SSN correlation to earlier decades}
% \section{Correlating NGK rY and sunspot numbers}
\label{sec:rY_earlier}
%%%%%%%%%%%%%%%%%%%%%%%%%%%%%%%%%%%%%%%%%%%%%%%%%%%%%%%

%%%%%%%%%% FIGURE 3 %%%%%%%%%%%%
  \begin{figure*}[h]  %% 
   \centering
  \includegraphics[width=0.98\linewidth]{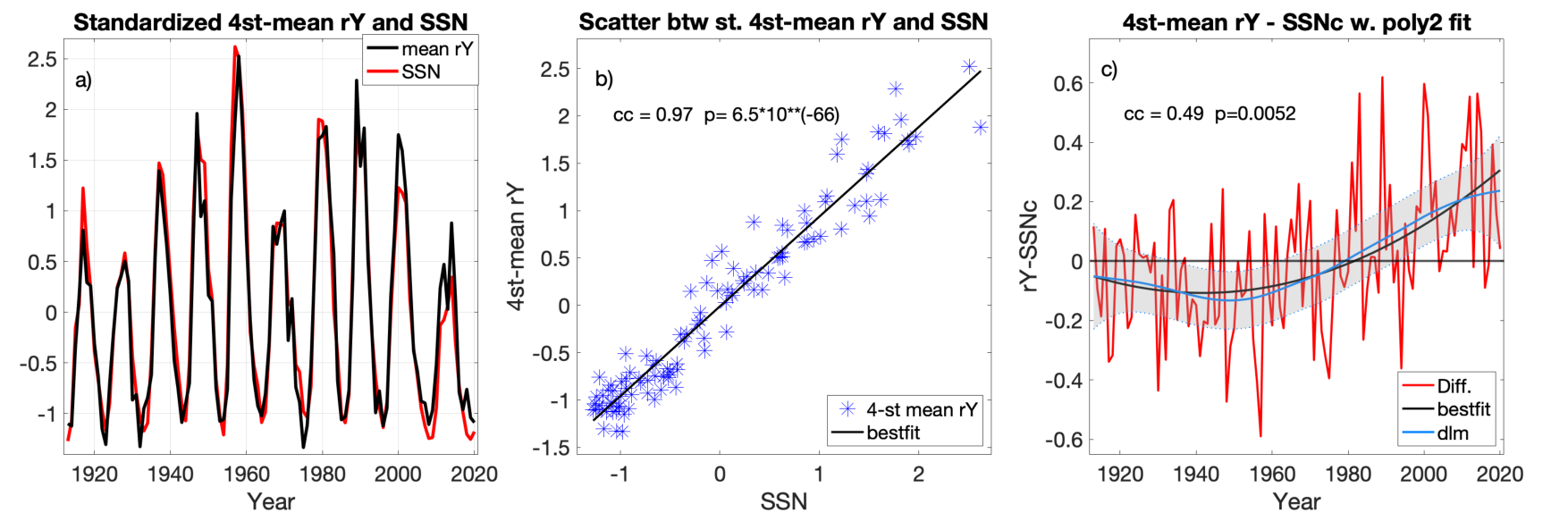}
  \includegraphics[width=0.98\linewidth]{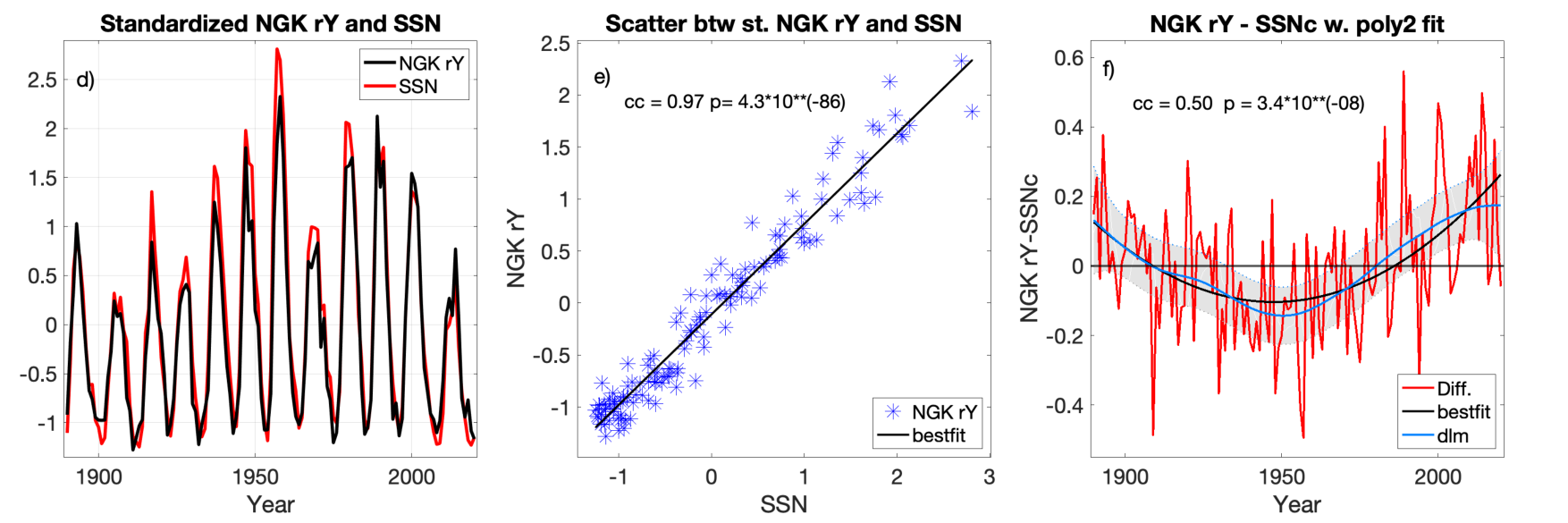}
    \caption{a) standardized yearly 4-station rY index (black) and sunspot number (red). 
    b) Scatterplot of standardized 4-st rY index and SSN (blue stars), the best-fit line (black) and its correlation coefficient and p-value. 
    c) Difference of standardized 4-st rY and correlated SSN (red).  
   Best-fit second-order polynomial (black) with correlation coefficient and p-value, and the dlm curve (blue) with 95\% errors around it (grey area). 
 d) - f) Same as panels a-c but for NGK rY index.  
   } 
   \label{fig:figure_rY_4st_mean_SSN_3panels}
\end{figure*}

%%%%%%%%%%%%%%%%%%%%%%%%%%%%%%%%%%%%%%%%%%
% 4-station mean rY 

We have extended the above analysis based on the 6-station mean to earlier decades in two steps.
First we used the mean of the four longest stations (HON, KAK, NGK, TUC) to extend the analysis by two decades to start in 1913.
Then we used the longest station (NGK) to extend the analysis by additional 20 years to cover a 130-year interval (1890-2020). %which includes the whole Modern Maximum.

We constructed the yearly 4-station (4-st) mean rY index in the same way as the 6-station mean described above.
Figure \ref{fig:figure_rY_4st_mean_SSN_3panels}a depicts the time series of the standardized 4-st mean rY index and the standardized sunspot number in 1913-2020.
Figure \ref{fig:figure_rY_4st_mean_SSN_3panels}a shows that the peaks of four of the first five SSN cycles are higher than the  rY cycle peaks, while the last three rY cycle peaks are higher than the corresponding SSN peaks.
These differences are well in line with Fig. \ref{fig:figure_rY_6st_mean_SSN_4panels}a, extending the SSN dominance to two earlier cycles.

Figure \ref{fig:figure_rY_4st_mean_SSN_3panels}b shows the scatterplot between the standardized 4-station mean rY index and the standardized sunspot number.
Despite a smaller number of station (and slightly larger scatter), the correlation is excellent (cc = 0.97), with sunspots explaining 
94\% of rY variation.
Although the correlation coefficient is slightly smaller than in Fig. \ref{fig:figure_rY_6st_mean_SSN_4panels}b, the p-value is six orders of magnitude smaller due to the longer time interval.

Figure \ref{fig:figure_rY_4st_mean_SSN_3panels}c depicts the difference (residual) between the standardized 4-station mean rY index and the correlated sunspot number obtained from the correlation of Fig. \ref{fig:figure_rY_4st_mean_SSN_3panels}b.
The nonlinear evolution of this difference is revealed by the additional decades included in the 4-station mean.
Due to an overall increasing trend, even a linear correlation (cc = 0.43) is significant, but a best-fit second order polynomial correlates slightly better (cc = 0.49).
Most interestingly, the coefficient of the second order term of this polynomial ($6.75\pm 2.36 \cdot 10^{-5}$) is nonzero at 99.4\% confidence level (p = 0.0052).
The dlm fit follows the second order polynomial closely, verifying the nonlinear evolution of the residual.

%%%%%%%%%%%%%%%%%%%%%%%%%%%%%%%%%%%%%%%%%%
% NGK  rY

Continuous observations at the NGK station 
started already in 1890, which allows to extend the rY index to even earlier times.
We have correlated the NGK rY index with the standardized 6-st mean rY index, 
thereby normalizing the NGK rY index to the same level and continuing the 6-st index to earlier decades (NGK rY index).
Figures \ref{fig:figure_rY_4st_mean_SSN_3panels}d-f depict the similar three panels for the NGK rY index as Figs. \ref{fig:figure_rY_4st_mean_SSN_3panels}a-c show for the 4-station mean rY index.
Figure \ref{fig:figure_rY_4st_mean_SSN_3panels}d shows the time series of the standardized NGK rY index and the standardized sunspot number in 1890-2020.
Again, sunspot cycles are higher than rY cycles when activity is increasing and high, and lower when it is decreasing during the last three cycles.
Interestingly, for cycle 13, the first cycle included in Fig. \ref{fig:figure_rY_4st_mean_SSN_3panels}d, the NGK rY cycle is higher than sunspot cycle.

Figure \ref{fig:figure_rY_4st_mean_SSN_3panels}e shows the scatterplot between the NGK rY index and sunspot number, whose %and the related best-fit line.
correlation coefficient (0.97) is as good as for the 4-station mean.
Moreover, the p-value ($4.3\cdot10^{-86}$) is 20 orders of magnitude smaller than in Fig. \ref{fig:figure_rY_4st_mean_SSN_3panels}b, again, due to the longer time series.
Figure \ref{fig:figure_rY_4st_mean_SSN_3panels}f depicts the difference between the standardized NGK rY index and the correlated sunspot number. 
This difference shows a clearly nonlinear evolution with a decreasing trend from 1890s until 1950s, and an increase from 1960s onward.
Figure \ref{fig:figure_rY_4st_mean_SSN_3panels}f also includes a best-fit second order polynomial which, despite the rather large yearly variability, explains a considerable fraction, about 25\% (cc = 0.50), of the variability in the difference.
Most importantly, the second-order term of this polynomial ($6.97\pm 1.19 \cdot 10^{-5}$) is very significantly nonzero (p = $3.4\cdot10^{-8}$).
This verifies the nonlinear temporal evolution of this difference.
On the other hand, a linear fit (not shown) is now hardly significant (p = 0.021) and explains only 4\% of variability.
The nonlinear evolution of the difference depicted in Fig. \ref{fig:figure_rY_4st_mean_SSN_3panels}f is also verified by the dlm fit.

These results give conclusive evidence for a temporally changing evolution of the rY - SSN relation during the last 130 years.
Note also that the dlm curve in Fig. \ref{fig:figure_rY_4st_mean_SSN_3panels}f seems to flatten out during the last two decades, slightly deviating from the second order evolution.
This may indicate that the long increase of the difference will soon change to a decrease, as solar activity has started increasing with cycle 25.

%%%%%%%%%%%%%%%%%%%%%%%%%%%%%%%%%%%%%%%%%%%%%%%%%%%%%%%
% DISCUSSION 
\section{Discussion}
\label{sec:Discussion}
%%%%%%%%%%%%%%%%%%%%%%%%%%%%%%%%%%%%%%%%%%%%%%%%%%%%%%%

We have used here the Earth's dayside ionospheric currents as a homogeneous long-term (centennial) proxy of solar EUV irradiance.
The same method was used, e.g., by \cite{Svalgaard_EUV_2016} to extend the solar EUV irradiance time series to earlier times.
Geomagnetic measurements are long-term homogeneous and experience no temporal degradation, contrary, e.g., to satellite EUV instruments that suffer from radiation damage caused by the measured EUV and other hard photons and solar energetic particles.
The current method is insensitive to solar flares which produce random deflections to the daily curve but average out in the yearly mean values.
Ground-based solar spectral line observations, such as chromospheric CaII K-line measurements, cover a century but suffer from changes in instrumentation, detection method, uneven sampling and in atmospheric transparency.
Accordingly, we are confident that the current method gives the only homogeneous estimate of EUV irradiance covering the whole Modern Maximum.

The Magnesium-II core-to-wing ratio measurements form the standard MgII index of solar EUV irradiance based on satellite solar observations since 1978 \citep[see, e.g.,][]{DeLand_1993, Viereck_1999, Viereck_2001, Viereck_2004, Snow_2014}.
%\citep[see, e.g.,][]{Tapping_2013}.
We have shown earlier 
that the MgII index increases relative to sunspot number from the end-1970s to 2010s \citep{Mursula_AA_2024}.
In addition, several proxies of solar EUV irradiance, like the solar radio fluxes F10.7, F15 and F30 have been found to increase with respect to the sunspot number from the 1960s to 2010s \citep{Bruevich_2019, Lastovicka_SpW_2023, Mursula_AA_2024, Tapping_2017}. 
Figure \ref{fig:S3} shows that, in agreement with these earlier results, there is no significant change in the relation between the MgII index and the 6-station mean rY index.
(Neither the linear (p=0.64) nor the quadratic (p=0.15) fit indicate a significant trend for the residual).
This verifies that the 6-st mean rY index is a reliable proxy of solar EUV irradiance during the last 40 years.

Here we have extended this earlier work and found that, during the low solar cycles at the turn of the 19th and 20th centuries, solar EUV irradiance dominated relative to sunspot numbers.
%, similarly as during the last few solar cycles. 
Thereafter, when solar cycle heights started increasing during the growth phase of the Modern Maximum, 
the sunspot numbers started dominating relative to solar EUV irradiance.
This changed again to EUV dominance when solar activity was reduced during the decline of the Modern Maximum.
This indicates a systematic variation in the mutual relation between sunspots and solar EUV irradiance during the Modern Maximum and, generally, during Gleissberg cycle -type large oscillations of overall solar activity.

Considering the recent efforts to improve the Wolf/International sunspot number series \citep{Clette_2015, Clette_Preface_2016}, one might suspect that this change in the relation between sunspots and EUV irradiance would be due to a possible inhomogeneity in the SSN.
However, the rY index has a similar centennially varying relation with the group sunspot number as with the SSN (see Fig. \ref{fig:S4}; the same result is also found for other versions of GSN).
This proves that our results are not an artefact due to a possible inhomogeneity in the SSN.

Solar EUV irradiance mainly originates in the chromosphere and lower corona above active regions called plages \citep{Pallavicini_1981, Neupert_1998}.
Accordingly, the found change in the long-term relation between sunspots and EUV irradiance suggests that the relation between photospheric sunspots and upper atmospheric (chromospheric and low coronal) plages changes during Gleissberg cycles.
Plages are rooted in the photosphere where they are seen as bright faculae. 
Although the long-term variation of faculae is less well known, it is commonly assumed that faculae follow the evolution of plages \citep[see, e.g.,][]{Yeo_2020}.
Faculae are important solar structures since the brightness (total solar irradiance, TSI) of the Sun and Sun-like stars is formed by the competing balance between darkening caused by sunspots and brightening produced by faculae \citep{Hall_2009, LockwoodGW_2007, Radick_2018}.

A fast-rotating young Sun created very strong magnetic fields with large sunspots which dominated brightness variation.
Later, in the current phase of solar stellar evolution, its magnetic fields are much weaker but still strong enough to create sunspots, although faculae now dominate brightness variation and produce the observed in-phase variation of TSI with sunspot cycle.
The Sun has relatively recently passed the transition from spot-dominated to facula-dominated phase in its stellar evolution \citep{Hall_2009, LockwoodGW_2007, Reinhold_2019}. 
Therefore, the balance between the two competing factors is particularly delicate for the current Sun.

Using a higher resolution (days to a few months) than here, the relation between faculae and sunspots is slightly nonlinear \citep{Foukal_1993, Foukal_1996, Chapman_1997}, reflecting the different decay times of sunspots and active regions \citep{Preminger_2005, Preminger_2007}.
This increases (decreases) the relative fraction of sunspots relative to faculae with increasing (decreasing, respectively) solar activity \citep{Foukal_1998, Reinhold_2019, Yeo_2020}.
Because of the temporally limited coverage of direct observations of solar chromospheric parameters, the relation between sunspots and faculae has so far been limited to the last 40-50 years, i.e., the time of decreasing solar activity.
Our results covering the whole Modern Maximum indicate that the sunspot-facula ratio is different during the growth and the decay phases of the Modern Maximum and, quite likely, during the corresponding phases of Gleissberg cycle -type solar and stellar activity variations. 
This difference will also modify the estimated evolution of solar/stellar brightness.

Concluding, we have used the range (amplitude) of the daily variation of the East (Y)-component as a homogenous centennial proxy of solar EUV irradiance to study the long-term relation between sunspot activity and solar EUV irradiance.
We have shown that sunspot activity dominated relative to EUV irradiance during increasing and high solar activity (during growth and maximum of the Modern Maximum, the latest Gleissberg cycle), while EUV irradiance dominated relative to sunspots during low cycles before the Modern Maximum and during the decay of the Modern Maximum. 
We showed that this change is not due to a possible small inhomogeneity of the sunspot number.
Since EUV irradiance mainly comes from plages/faculae, this suggests that the relation between spots and faculae changes (oscillates) during Gleissberg cycle -type large activity oscillations.
These results indicate so far unknown, activity-related long-term changes in the solar atmosphere that have relevance to the long-term evolution of the spot-facula ratio and brightness of the Sun and Sun-like stars.

%%%%%%%%%%%%%%%%%%%%%%%%%%%%%%%%%%%%%%%%%%%%%%%%%%%%%%%%
%% CONCLUSIONS
%\section{Conclusions}
%\label{Sec: Conclusions}

%%%%%%%%%%%%%%%%%%%%%%%%%%%%%%%%%%%%%%%%%%%%%%%
% ACKNOWLEDGEMENTS

\begin{acknowledgements}

The magnetic data were provided by the World Data Center for Geomagnetism at the British Geological Survey (https://wdc-dataportal.bgs.ac.uk). 
The sunspot numbers (version 2; https://www.sidc.be/SILSO/datafiles) and the revised collection of the number of sunspot groups (https://www.sidc.be/SILSO/groupnumberv3) were provided by the World Data Center SILSO.
Dr. Mark A. Weber is acknowledged for constructing and maintaining the web site for the Bremen MgII composite index.
The dynamic linear model of Marko Laine can be found at https://mjlaine.github.io/dlm/index.html.

\end{acknowledgements}

%%%%%%%%%%%%%%%%%%%%%%%%%%%%%%%%%%%%%%%%%%%%%%%
%% REFERENCES

% Please note that we have included the references below in
% order to compile the document, but we ask you to:

\bibliographystyle{aa} % style aa.bst

\bibliography{rY_SSN_FINAL}

%%%%%%%%%%%%%%%%%%%%%%%%%%%%%%%%%%%%%%%%%%%%%%%
% APPENDICES
%
%%%%%%%%%%%%%%%%%%%%%%%%%%%%%%%%%%%%%%%%%%%%%%%%%%%%%%%%%%%%%%%
% Appendices must be placed after   \end{thebibliography}
% They will be placed automatically on a new page.
%%%%%%%%%%%%%%%%%%%%%%%%%%%%%%%%%%%%%%%%%%%%%%%%%%%%%%%%%%%%%%%
\begin{appendix}
%%%%%%%%%%%%%%%%%%%%%%%%%%%%%%%%%%%%%%%%%%%%%%%%%%%%%%%%%%%%%%%
% In the PDF output, floats should be placed
% under their own appendix, not before the title, nor after the
% title of the next appendix.

% In short appendices, onecolumn floats (\figure*
% or \table*) will generate a blank page.
% To prevent this behaviour, a few examples are provided here. 

% In case you have a lot of floating objects for little text and the 
% LaTeX engine moves the floats away from their context, the command
% \FloatBarrier of the “placeins” package will empty the
% float buffer and place all stored floats in the continuity.

% If you still encounter problems with wide floats placement,
% just use the onecolumn environment throughout the appendices.
%%%%%%%%%%%%%%%%%%%%%%%%%%%%%%%%%%%%%%%%%%%%%%%%%%%%%%%%%%%%%%%

%%%%%%%%%%%%%%%%%%%%%%%%%%%%%%%%%%%%%%%%%%%%%%%%%%%%%%%%%%%%%%%
%  APPENDIX A. 

\onecolumn
\section{Stations and daily variation of Y-component}

Table \ref{table:Stations} shows the IAGA abbreviations, the geographic latitudes and longitudes, the UT hour of midnight, and the start year of operations of the six magnetic stations used in this study: Hermanus (HER), Honolulu (HON), Kakioka (KAK), Niemegk (NGK), San Juan (SJG) and Tucson (TUC).

\begin{figure*}[h!]  %%  Globe with 2 stations LT curves
   \centering
  \includegraphics[width=0.4\linewidth]{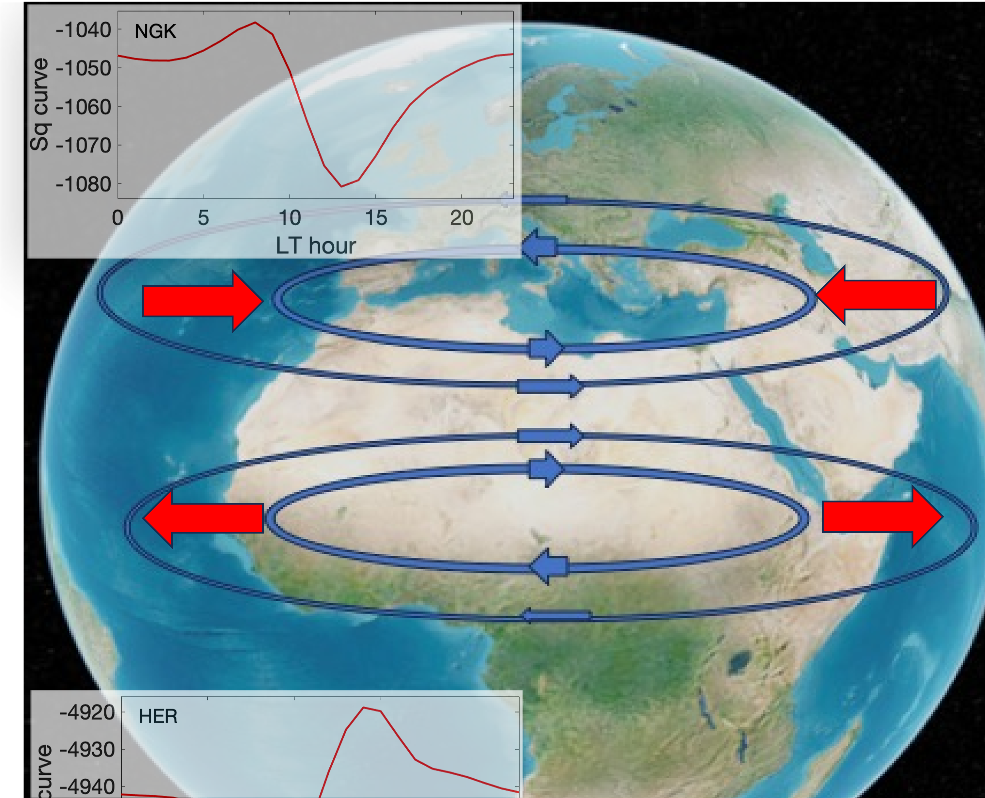}
  \includegraphics[width=0.55\linewidth]{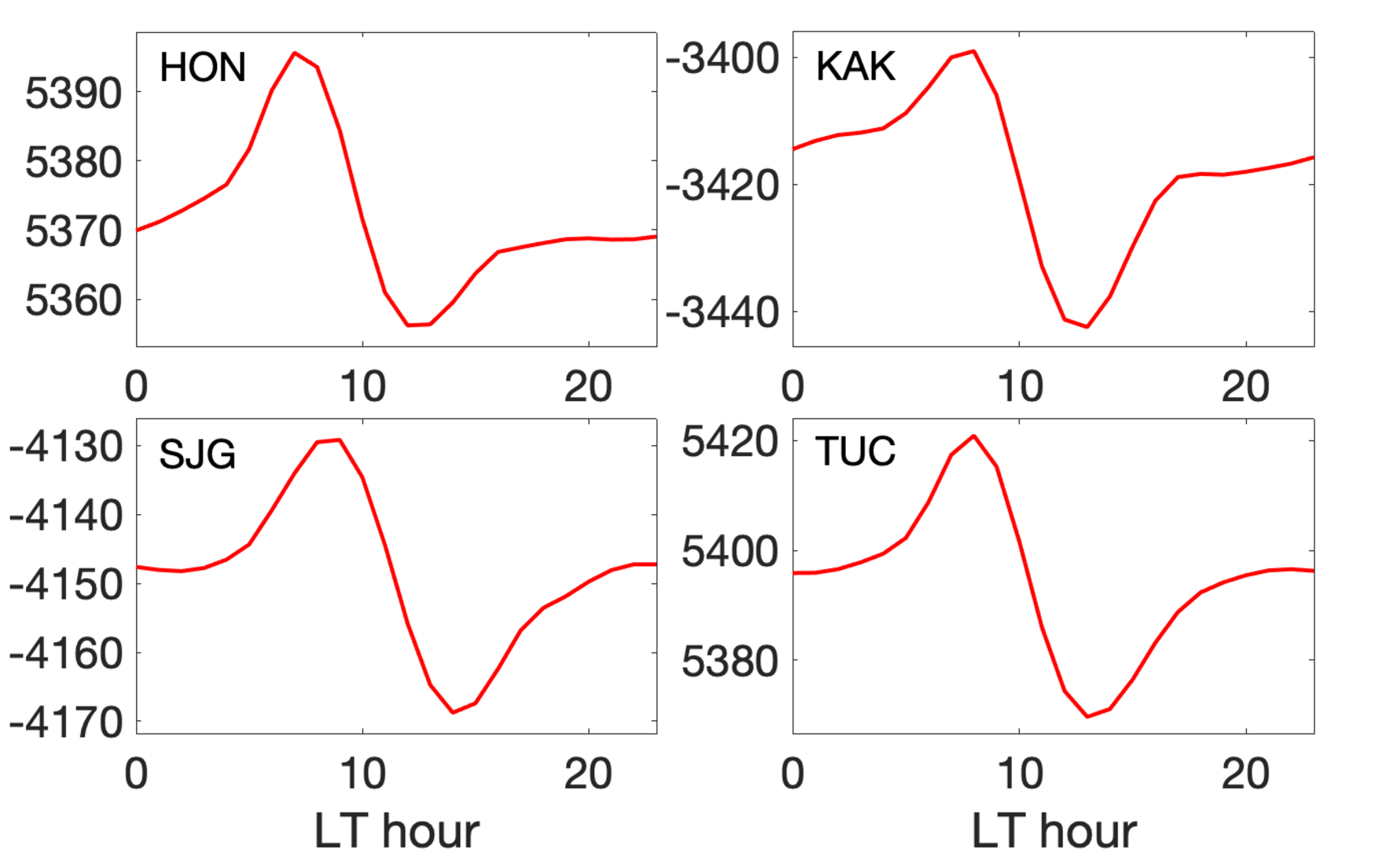}
    \caption{
 Left panel: Sketch of the $S_q$-current system (blue lines with arrows). Red thick arrows denote the magnetic effect of the $S_q$-current on the Y-component, whose mean daily curves (thin red line) for NGK and HER are depicted in the two insets of the panel. Right panel: Daily curves of the Y-component for the other four stations. 
  } 
 \label{fig:S1}
\end{figure*}

The left panel of Fig. \ref{fig:S1} depicts a sketch of the $S_q$-current system (blue lines with arrows) at 11 UT when the subsolar region is at the longitude of NGK and HER. 
Red thick arrows denote the magnetic effect of the $S_q$-current on the Y-component, whose mean daily curves for NGK and HER are depicted in the two insets of this panel. 
The right panel of Fig.  \ref{fig:S1} depicts the mean daily curves of the Y-component for the other four stations. 
Note that, although the daily curves are very similar, their absolute level can be positive or negative, depending on the location of the station. 
For the secular change of the Y-component and the yearly maxima and minima of the daily curve at the six stations, see Fig. \ref{fig:S2}.

%%%%%%%%%% TABLE 1  %%%%%%%%%%%%
\begin{table*}[h!]
\caption{Abbreviations, geographic latitudes and longitudes, UT hour of midnight, and start year of the six magnetic stations.}             % \caption{Magnetic stations used: abbreviations, geographic latitudes and longitudes, UT hour of midnight, start year of operations.}             % title of Table
\label{table:Stations}   % is used to refer to this table in the text
\centering                          		% used for centering table
\begin{tabular}{|clc|clc|cl}			% centered columns
\hline					 % inserts single horizontal line
&\textbf{GGlat}&\textbf{GGLong}&\textbf{MidnightUT}&\textbf{Start year}&\\
\hline % table heading
\textbf{HER}&-34.425&19.225&23&1932&\\\hline
\textbf{HON}&21.317&-158&11&1902&\\\hline
\textbf{KAK}&36.233&140.183&15&1913&\\\hline
\textbf{NGK}&52.072&12.675&23&1890&\\\hline
\textbf{SJG}&18.382&-66.118&4&1926&\\\hline
\textbf{TUC}&32.17&-110.73&7&1909&\\\hline
\end{tabular}
\end{table*}

%%%%%%%%%% FIGURE S1 %%%%%%%%%%%%
  
% \clearpage

%%%%%%%%%%%%%%%%%%%%%%%%%%%%%%%%%%%%%%%%%%%%%%%%%%%%%%%%%%%%%%%
%  APPENDIX B 

\section{Long-term variation of Y-component and daily minimum and maximum}

\begin{figure*}[h!] 		%% 6 stations secular curves
%   \centering
%  \includegraphics[width=0.98\linewidth]{worldmap_new_6st_924.pdf}
\noindent\includegraphics[width=\textwidth]{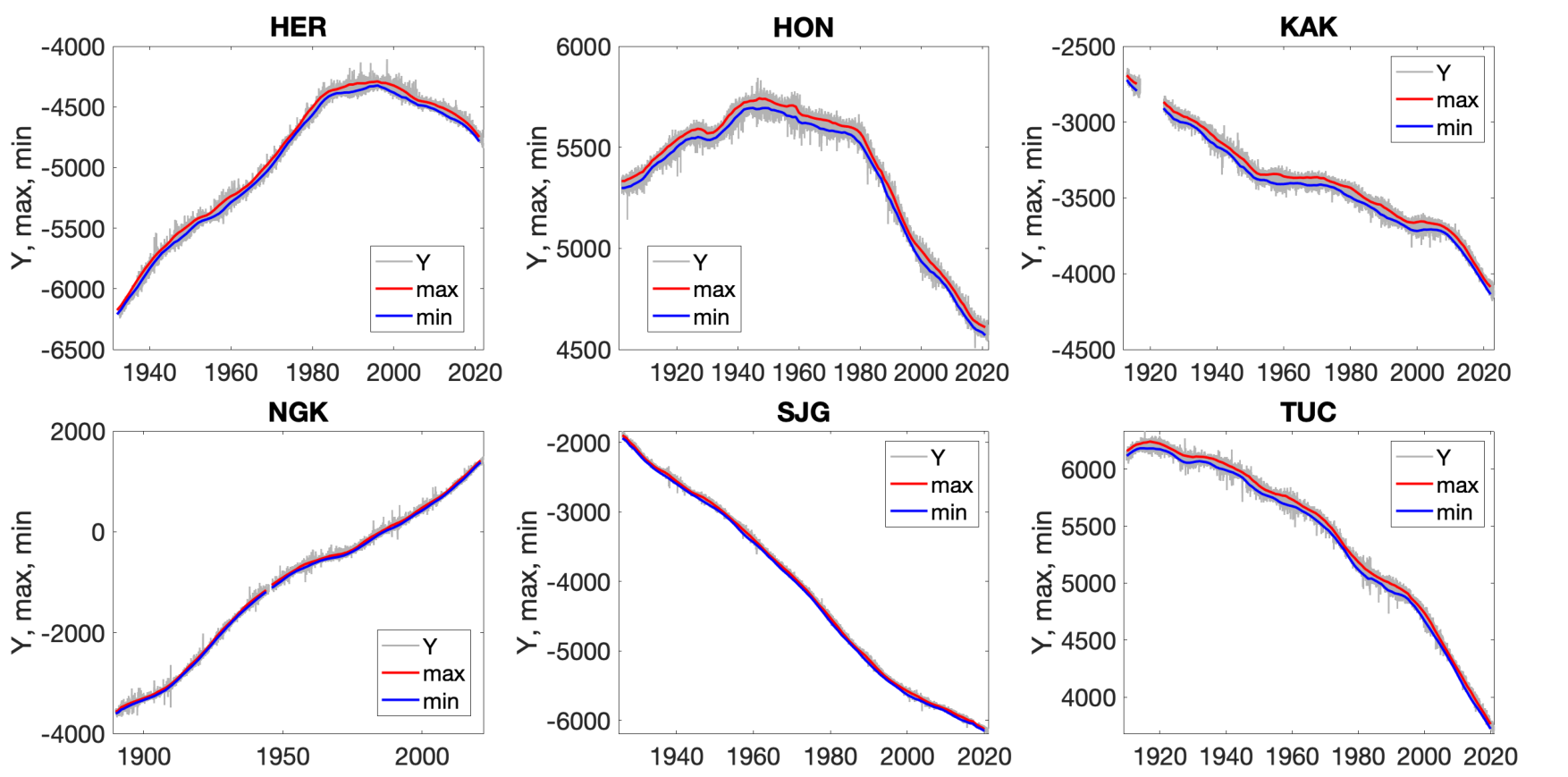}
    \caption{
Secular variation of the Y-component of the six stations at hourly resolution (grey line), as well as the yearly means of the daily maxima (red curve) and minima (blue curve) of the Y-component of each station. The unit of the y-axis is nanotesla (nT). 
 } 
 \label{fig:S2}
%\end{figure*}
\end{figure*}

The six panels of Fig. \ref{fig:S2} depict the long-term (secular) variation of the Y-component of the six stations at hourly resolution (grey line), as well as the yearly means of the daily maxima (red curve) and minima (blue curve) of the Y-component of each station. 
The form, scale and the range of variation of the Y-component of the six stations are very different, depending on the location of the station. 
The range of secular variation varies from the largest range of more than 4000\,nT at SJG to the smallest range of about 1200\,nT at HON. 
Note that, since the mean difference of the daily maximum (red) and minimum (blue) is roughly the same, about 30\,nT for each station (see Fig. \ref{fig:figure_rY_6st_abs_st_mean}), the separation (and visibility) of the two colored curves in Fig. \ref{fig:S2} varies with the range of secular variation.

%%%%%%%%%% FIGURE S2 %%%%%%%%%%%%
%  \begin{figure*}[h]  

%%%%%%%%%%%%%%%%%%%%%%%%%%%%%%%%%%%%%%%%%%%%%%%%%%%%%%%%%%%%%%%
%  APPENDIX C 

\onecolumn

\section{Relation between solar MgII index and 6-station mean rY index}

\begin{figure*}[h!] 		%% 6 stations secular curves
%   \centering
% \includegraphics[width=0.98\linewidth]{figure_NGK_rY_GSNVaq_3panels.pdf}
\noindent\includegraphics[width=\textwidth]{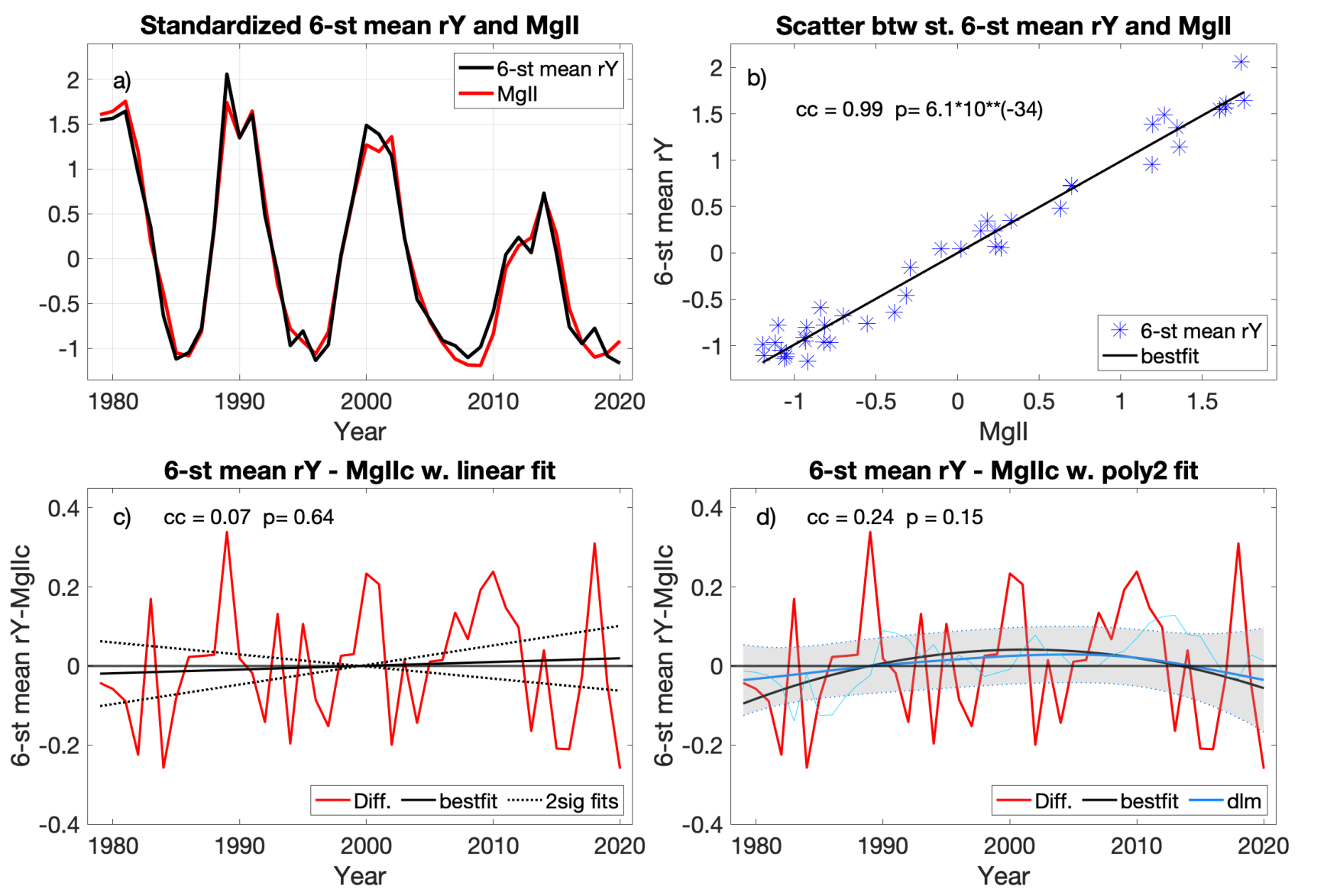}
    \caption{
    a) standardized 6-station mean rY index (black) and MgII index (red). 
    b) Scatterplot of standardized 6-st mean rY index and MgII index (blue stars), the best-fit line (black) and its correlation coefficient and p-value. 
    c) Difference of standardized 6-st mean rY and correlated MgII index.  
   Best-fit line (solid) with correlation coefficient and p-value, as well as lines with slopes that are two standard deviations above or below the best-fit line slope (dotted) are also included. 
   d) Same difference of standardized 6-st mean rY and correlated MgII index as in c), with best-fit second-order polynomial fit (black) and the dynamic linear model curve (blue) with 95\% errors (grey area).  
   } 
 \label{fig:S3}
%\end{figure*}
\end{figure*}

Figure \ref{fig:S3} depicts the relation between the standardized yearly 6-station mean rY index and the MgII index in 1979-2020 in the same format as the four panels of Fig. \ref{fig:figure_rY_6st_mean_SSN_4panels} for the 6-station mean rY index and sunspot number. 
Figure \ref{fig:S3}a shows that the cycle peaks or minima of the two parameters do not depict the same systematic tendencies as seen in Fig. \ref{fig:figure_rY_6st_mean_SSN_4panels}. 
Their mutual correlation (cc = 0.99) is excellent (see Fig. \ref{fig:S3}b). 
The difference (residual) between the rY index and the correlated MgII index does not depict any significant trend for neither the linear fit (p=0.64; Fig. \ref{fig:S3}c) nor for the quadratic fit (p=0.15; Fig. \ref{fig:S3}d). 
The slope of the best-fit line is practically zero and the coefficient of the second order term is zero within the standard 95\% confidence limit. 
Moreover, the second oder polynomial always stays within the 95\% confidence limit of the dlm fit.

 %%%%%%%%%% FIGURE S3 %%%%%%%%%%%%
%  \begin{figure*}[h]  

%%%%%%%%%%%%%%%%%%%%%%%%%%%%%%%%%%%%%%%%%%%%%%%%%%%%%%%%%%%%%%%
%  APPENDIX D 

 \section{Relation between GSN and NGK rY index}

\begin{figure*}[h!]  		%% 6 stations secular curves
%   \centering
% \includegraphics[width=0.98\linewidth]{figure_NGK_rY_GSNVaq_3panels.pdf}
\noindent\includegraphics[width=\textwidth]{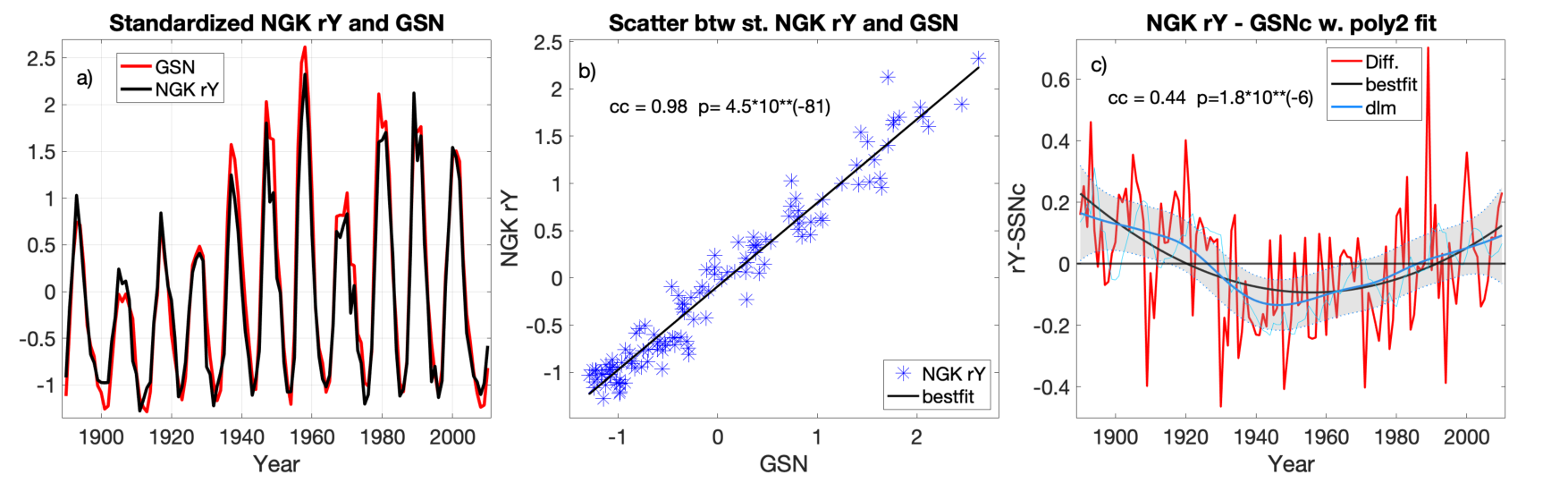}
    \caption{
    a) standardized yearly NGK rY index (black) and group sunspot number (red). 
    b) Scatterplot of standardized NGK rY index and GSN (blue stars), the best-fit line (black) and its correlation coefficient and p-value. 
    c) Difference of standardized NGK rY and correlated GSN (red).  
   Best-fit second-order polynomial (black) with correlation coefficient and p-value, and the dlm curve (blue) with 95\% errors around it (grey area).
   } 
 \label{fig:S4}
%\end{figure*}
\end{figure*}

Figure \ref{fig:S4} depicts the relation between the standardized yearly NGK rY index and the group sunspot number (GSN) \citep{Vaquero_2016} in 1890-2010 in the same format as panels d-f of Figure \ref{fig:figure_rY_4st_mean_SSN_3panels} for the NGK rY and Wolf/International sunspot number (SSN) in 1890-2020. 
(Note that the GSN series misses the last cycle 24).
Panel a of Fig. \ref{fig:S4} shows that the NGK rY cycle peaks are higher than GSN peaks for the first three cycles and the last two cycles, while the reverse order applies for the remaining cycles. 
These relations are very similar as those between NGK rY and SSN, as seen in Fig. \ref{fig:figure_rY_4st_mean_SSN_3panels}d. 
The correlation between NGK rY and GSN (cc = 0.98) is excellent (see panel b), although the p-value is five orders of magnitude larger than in Fig. \ref{fig:figure_rY_4st_mean_SSN_3panels}e because of a shorter time series. 

The difference (panel c) between the NGK rY index and the correlated GSN depicts a significant quadratic relation. 
The correlation coefficient (cc = 0.44) is only slightly smaller and the p-value ($1.8\cdot 10^{-6}$) somewhat larger than for the correlation between NGK rY and SSN.

%%%%%%%%%% FIGURE S4 %%%%%%%%%%%%
%  \begin{figure*}[h]  

\end{appendix}
\end{document}